\documentclass[
 superscriptaddress,
reprint,
showpacs,
 bibnotes,
 showkeys,
 amsmath,amssymb,
 aps,
]{revtex4-1}
\usepackage{slashbox,pict2e}
\usepackage{graphicx}
\usepackage{amsmath}
\usepackage{dcolumn}
\usepackage{bm}
\usepackage{braket} 
\usepackage{bbold}
\usepackage[
 colorlinks=true,
 urlcolor=blue,
 citecolor=blue,
 linkcolor=blue,
 bookmarks=false,
 pdfstartview={FitH},
]{hyperref}

\pdfoutput=1

\def\KCA{K$_2$Cr$_3$As$_3$\,}

\begin{document}

\title{Observation of Raman active phonon with Fano lineshape in quasi-one-dimensional superconductor \KCA}
\author{W.-L. Zhang}
\thanks{These two authors contributed equally to this work.}
\affiliation{Beijing National Laboratory for Condensed Matter Physics and Institute of Physics, Chinese Academy of Sciences, Beijing, 100190, China} 
\author{H. Li}\thanks{These two authors contributed equally to this work.}
\affiliation{Beijing National Laboratory for Condensed Matter Physics and Institute of Physics, Chinese Academy of Sciences, Beijing, 100190, China} 
\author{Dai Xia}
 \affiliation{Beijing National Laboratory for Condensed Matter Physics and Institute of Physics, Chinese Academy of Sciences, Beijing, 100190, China}
 \author{H. W. Liu}
 \affiliation{Beijing National Laboratory for Condensed Matter Physics and Institute of Physics, Chinese Academy of Sciences, Beijing, 100190, China}
 \author{Y.-G. Shi}
\affiliation{Beijing National Laboratory for Condensed Matter Physics and Institute of Physics, Chinese Academy of Sciences, Beijing, 100190, China}
 \author{J. L. Luo}
\affiliation{Beijing National Laboratory for Condensed Matter Physics and Institute of Physics, Chinese Academy of Sciences, Beijing, 100190, China}
 \affiliation{Collaborative Innovation Center of Quantum Matter, Beijing, China}
 \author{Jiangping Hu}
 \affiliation{Beijing National Laboratory for Condensed Matter Physics and Institute of Physics, Chinese Academy of Sciences, Beijing, 100190, China}
 \affiliation{Collaborative Innovation Center of Quantum Matter, Beijing, China}
 \affiliation{Department of Physics, Purdue University, West Lafayette 47907, USA}
\author{P. Richard}
 \email{p.richard@iphy.ac.cn}
 \affiliation{Beijing National Laboratory for Condensed Matter Physics and Institute of Physics, Chinese Academy of Sciences, Beijing, 100190, China} 
 \affiliation{Collaborative Innovation Center of Quantum Matter, Beijing, China}
\author{H. Ding}
 \affiliation{Beijing National Laboratory for Condensed Matter Physics and Institute of Physics, Chinese Academy of Sciences, Beijing, 100190, China} 
 \affiliation{Collaborative Innovation Center of Quantum Matter, Beijing, China}
 
\date{\today}


\begin{abstract}
We performed a polarized Raman scattering study of quasi-one-dimensional superconductor K$_2$Cr$_3$As$_3$. We detect two A$_1^\prime$ phonons and three E$^\prime$ phonons. One of the A$_1^\prime$ modes exhibits a nearly temperature-independent Fano lineshape. Based on our first-principles calculations, we ascribe this mode to the in-phase vibrations of Cr atoms within one layer. This observation strongly suggests that the magnetic fluctuations in K$_2$Cr$_3$As$_3$ are coupled to the electronic structure \emph{via} the lattice.
\end{abstract}


\maketitle

Superconductivity in conventional materials is mediated by the electron-phonon interaction. When the electron-phonon coupling is strong enough, the spectral lineshape of Raman phonon excitations is characterized by an asymmetric profile called Fano lineshape \cite{Fano_PhysRev1961}. Despite a superconducting transition temperature ($T_c$) up to 6.1~K comparable to that of conventional superconductors, the recently discovered quasi-one-dimensional superconductors A$_2$Cr$_3$As$_3$ (A = K, Rb, Cs) have exotic properties that tentatively earned them the label of unconventional superconductors~\cite{JKBao_PRX2015,ZTTang_SCM2015,ZTTang_PRB2015}, such as an abnormal power law divergence of the $^{75}$As nuclear spin lattice relaxation rate near $T_c$~\cite{HZZhi_NMR2015}. Penetration depth measurements suggest a nodal line in the superconducting gap~\cite{GMPang_arXiv2015} and a large zero temperature upper critical field $H_{c2}(0)$ compared to the Pauli limit has been attributed to a spin triplet $p$-wave pairing~\cite{JKBao_PRX2015}, in agreement with a theoretical study revealing strong ferromagnetic fluctuations caused by the frustration of Cr moments within one layer~\cite{XXWu_arxiV1503}. Interestingly, CrAs itself exhibits superconductivity in proximity to a magnetic instability~\cite{W_Wu5ncomm}. Moreover, theoretical calculations of the electronic density-of-states indicate a sizable renormalization factor in K$_2$Cr$_3$As$_3$ compatible with a large specific heat coefficient~\cite{JKBao_PRX2015}, suggesting significant electron correlations~\cite{HaoJiang_arXiv2014,XXWu_CPL2015}. However, although some evidences push towards unconventional superconductivity in K$_2$Cr$_3$As$_3$, this view is not universally accepted ~\cite{KTai_PRB2015} and whether the electron-phonon interaction plays a key role in this material remains an open question.

\begin{figure}[!t]
\begin{center}
\includegraphics[width=0.9\columnwidth]{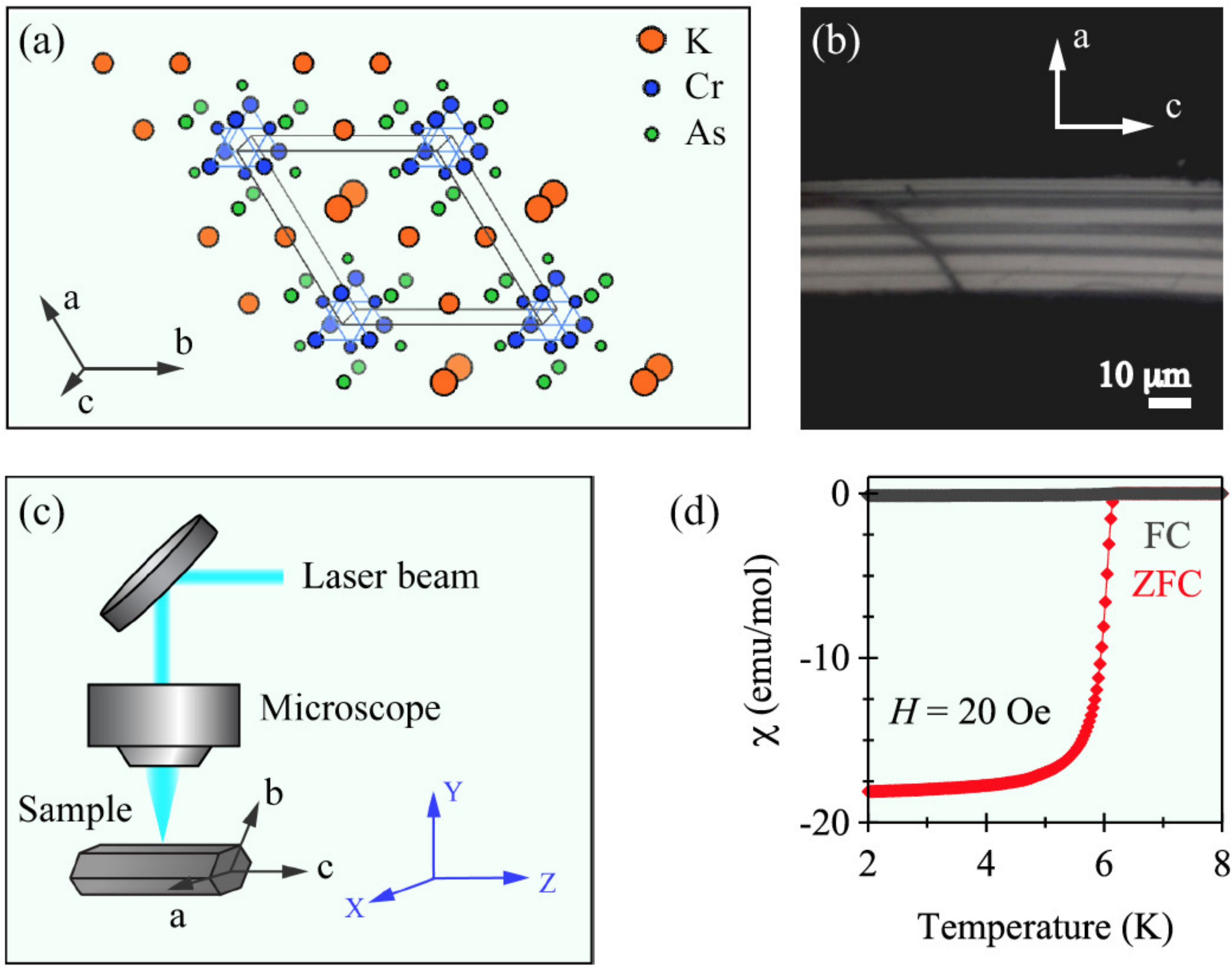}
\end{center}
 \caption{\label{Fig: 1} (Color online) (a) Unit cell of K$_2$Cr$_3$As$_3$. (b) Microscope photo of the ac surface of a measured \KCA sample. 
(c) Optical setup and choice of crystallographic and laboratory coordinates. (d) Magnetic susceptibility of one sample from the same batch.}
\end{figure}

In this Letter, we report a polarized Raman scattering study of K$_2$Cr$_3$As$_3$. We identify two and three phonon modes with the A$_1^\prime$ and E$^\prime$ symmetries, respectively. Surprisingly, one of the A$_1^\prime$ phonons exhibits a Fano lineshape, nearly temperature-independent from 6 K to room temperature, which is characterized by a coupling strength factor $1/q = - 0.29$. Our theoretical analysis indicates that this mode involves the in-plane movement of the Cr atoms, which directly modulates the Cr-Cr bonding identified to be critical to the magnetic fluctuations in this system. Although we cannot rule out a direct electron-phonon coupling in K$_2$Cr$_3$As$_3$, our results are rather suggestive of a lattice-mediated coupling between electrons and magnetic fluctuations. 

The \KCA single crystals used in this Raman study have been synthesized by a self-flux method~\cite{JKBao_PRX2015}. The needle-like samples have a typical size of 0.1$\times$0.1$\times$3~mm$^3$. The crystal structure belongs to the space group P$\bar{6}$m2 (point group D$_{3h}$), with the needle direction coinciding with the c axis (Figs.~\ref{Fig: 1}(a)-\ref{Fig: 1}(c)). Magnetic susceptibility measurements indicate that $T_c$ is 6.1~K (Fig.~\ref{Fig: 1}(d)).
The samples were prepared and cleaved in Ar or N$_2$ atmosphere and moved into vacuum conditions without any exposure to air. The freshly cleaved single crystals have been measured in a back-scattering geometry from the ac and bc surfaces (Figs.~\ref{Fig: 1}(b) and \ref{Fig: 1}(c)). We used the 488 and 514~nm lines of Ar-Kr ion laser to focus a spot smaller than 5$\times$5~$\mu$m$^2$ by using a microscope with the total incident power less than 0.25~mW. The Raman scattering signal was analyzed by a Horiba Jobin Yvon T64000 spectrometer and collected by a liquid N$_2$ cooled CCD detector. The data were corrected for the CCD dark current and the luminescence background determined by extrapolating the featureless linear spectra between 400 to 750~cm$^{-1}$. The temperature dependent measurements from 6~K to 300~K were performed in a Janis liquid helium flow cryostat in a working vacuum better than 2$\times$10$^{-6}$~mbar. The Raman spectra were obtained in all informative back-scattering configurations, which correspond to $e^ie^s$ = XX, ZZ and XZ. Here $e^i$ and $e^s$ are the polarizations of the incident and scattered light defined in the laboratory coordinates, with X and Z corresponding to the crystallographic axes a and c, respectively, and Y oriented at 30 degrees from the b axis (Fig.~\ref{Fig: 1}(c)). 
 
We also performed first-principles calculations using the projector augmented wave (PAW) method encoded in the Vienna \emph{ab initio} simulation package (VASP) \cite{Kresse1993,Kresse1996,Kresse1996B} and the generalized-gradient approximation (GGA) \cite{Perdew1996} for the exchange correlation functions. The cutoff energy was set to be 500 eV and the number of $k$ points was set to $2\times2\times5$ for a $2\times2\times2$ supercell. The real-space force constants of the supercells were calculated in the density-functional perturbation theory (DFPT) \cite{Baroni} and the phonon frequencies were calculated from the force constants using the PHONOPY code \cite{phonopy}. As the Cr-Cr bond lengths are underestimated after relaxation of the lattice~\cite{XXWu_CPL2015}, which may be related to the strong spin fluctuations in this material, all phonon modes were calculated using the experimental lattice parameters~\cite{JKBao_PRX2015}.

\begin{figure}[!t]
\begin{center}
\includegraphics[width=1.0\columnwidth]{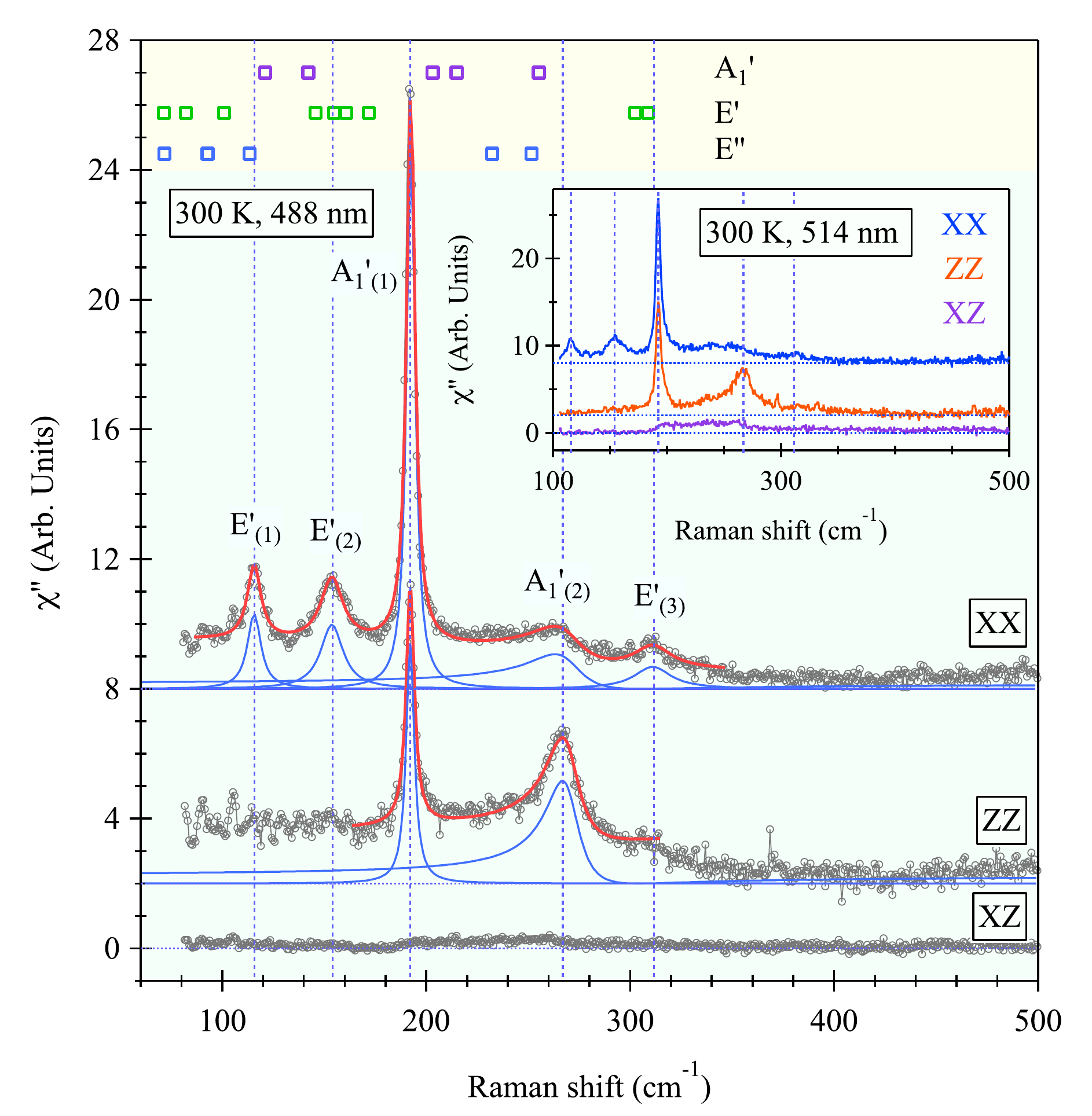}
\end{center}
 \caption{\label{Fig: 2} (Color online) Raman susceptibility at 300 K measured under different polarization configurations with 488 nm laser excitation. Each peak can be fitted by a Lorentz or a Fano function for the spectra under XX and ZZ polarizations. The individual peak functions are represented by blue curves while the red curves represent the resulting fitted spectra. The squares at the panel top represent the calculated BZ center phonon energies (see Table \ref{Table: Cal}). From top to bottom are the calculated results for the A$^\prime$, E$^\prime$ and E$^{\prime\prime}$ modes, respectively. Inset: Raman susceptibility at 300 K under different polarizations with 514 nm laser excitation.}
\end{figure}

With two formulas in one unit cell (\emph{i.e.} 16 atoms), K$_2$Cr$_3$As$_3$ is characterized by 48 phonon vibrational modes at the Brillouin zone (BZ) center ($\Gamma$ point). A group theory analysis shows that the corresponding irreducible representations are: [A$_2^{\prime\prime}$ + E$^\prime$] + [5 A$_1^\prime$ + 10 E$^{\prime}$ + 5 E$^{\prime\prime}$] + [5 A$_2^{\prime}$ + 5 A$_2^{\prime\prime}$], where the first, second and third terms are the acoustic, Raman active and infrared active modes, respectively. The corresponding Raman tensors in the laboratory coordinates are~\cite{Bilbao_1}: 
\[ R_{A_1^\prime} = \left( \begin{array}{ccc}
a & 0 & 0 \\
0 & a & 0\\
0 & 0 & b \end{array} \right), \]\,\,
\[ \left\{R_{E^\prime_X} = \left( \begin{array}{ccc}
d & 0 & 0 \\
0 & -d & 0\\
0 & 0 & 0 \end{array} \right),\,\,
R_{E^{\prime}_Y} = \left( \begin{array}{ccc}
0 & -d &  0\\
-d & 0 & 0\\
0 & 0 & 0 \end{array} \right)\right\},\]
\[\left\{R_{E^{\prime\prime}_X} = \left( \begin{array}{ccc}
0 & 0 & c \\
0 & 0 & 0\\
c & 0 & 0 \end{array} \right),\,\,
R_{E^{\prime\prime}_Y} = \left( \begin{array}{ccc}
0 & 0 &  0\\
0 & 0 & -c\\
0 & -c & 0 \end{array} \right)\right\},\]\,\,

\noindent Here $a$, $b$, $c$ and $d$ refer to Raman tensor elements. The XX polarization can couple to A$_1^\prime$ and E$^\prime$ modes, whereas the ZZ and XZ polarizations only couple to A$_1^\prime$ and E$^{\prime\prime}$ modes, respectively. 

\begin{table}[h] 
\caption{Raman active phonon energy at $\Gamma$ by calculation and experiment at 300 K and the atoms primarily involved.} 
\label{Table: Cal}
\centering 
\begin{tabular}{ccccc} 
\hline\hline 
No. &Sym.& Cal. (cm$^{-1}$)& Exp. (cm$^{-1}$)&Main atoms involved\\ 
\hline 
1&$E^\prime$&30.648  &-&Cr$_1$,Cr$_2$\\ 
2&$E^\prime$& 71.304 &-&K$_1$,Cr$_2$,As$_1$\\
3&$E^{\prime\prime}$&71.528  &-&K$_1$\\
4&$E^\prime$& 82.203&-&K$_1$,K$_2$,Cr$_1$,As$_2$\\
5&$E^{\prime\prime}$&92.773&-&K$_1$,Cr$_2$,As$_2$\\
6&$E^\prime$& 100.731 & 115.7&K$_1$,K$_2$\\
7&$E^{\prime\prime}$& 113.377 &-&K$_1$,Cr$_1$,As$_1$\\
8&$A_1^\prime$&120.945 &-&K$_1$\\
9&$A_1^\prime$&142.218  &-&K$_1$,Cr$_1$,Cr$_2$\\
10&$E^\prime$&  145.631&-&K$_1$,K$_2$,As$_1$\\
11&$E^\prime$& 154.749 &154.1&K$_1$,Cr$_2$,As$_1$\\
12&$E^\prime$&  160.606&-&As$_1$,As$_2$\\
13&$E^\prime$&  171.732&-&K$_1$,Cr$_1$,Cr$_2$\\
14&$A_1^\prime$&203.121&192.1&As$_1$,As$_2$\\
15&$A_1^\prime$& 214.914 &-&Cr$_1$,Cr$_2$,As$_1$,As$_2$\\
16&$E^{\prime\prime}$& 232.367 &-&Cr$_2$\\
17&$E^{\prime\prime}$& 251.674 &-&Cr$_1$\\
18&$A_1^\prime$&255.194  & 267.0&Cr$_1$,Cr$_2$\\
19&$E^\prime$&302.336 &-&Cr$_1$,As$_2$\\
20&$E^\prime$&308.602 &311.0&Cr$_2$,As$_1$\\
\hline
\hline
\end{tabular} 
\label{table:symmetry} 
\leftline{The subscript numbers of atoms refer to the two equivalent }
\leftline{layers, as shown in Fig.~\ref{Fig: 3}(g).}
\end{table}

\begin{figure*}[!t]
\begin{center}
\includegraphics[width=2.0\columnwidth]{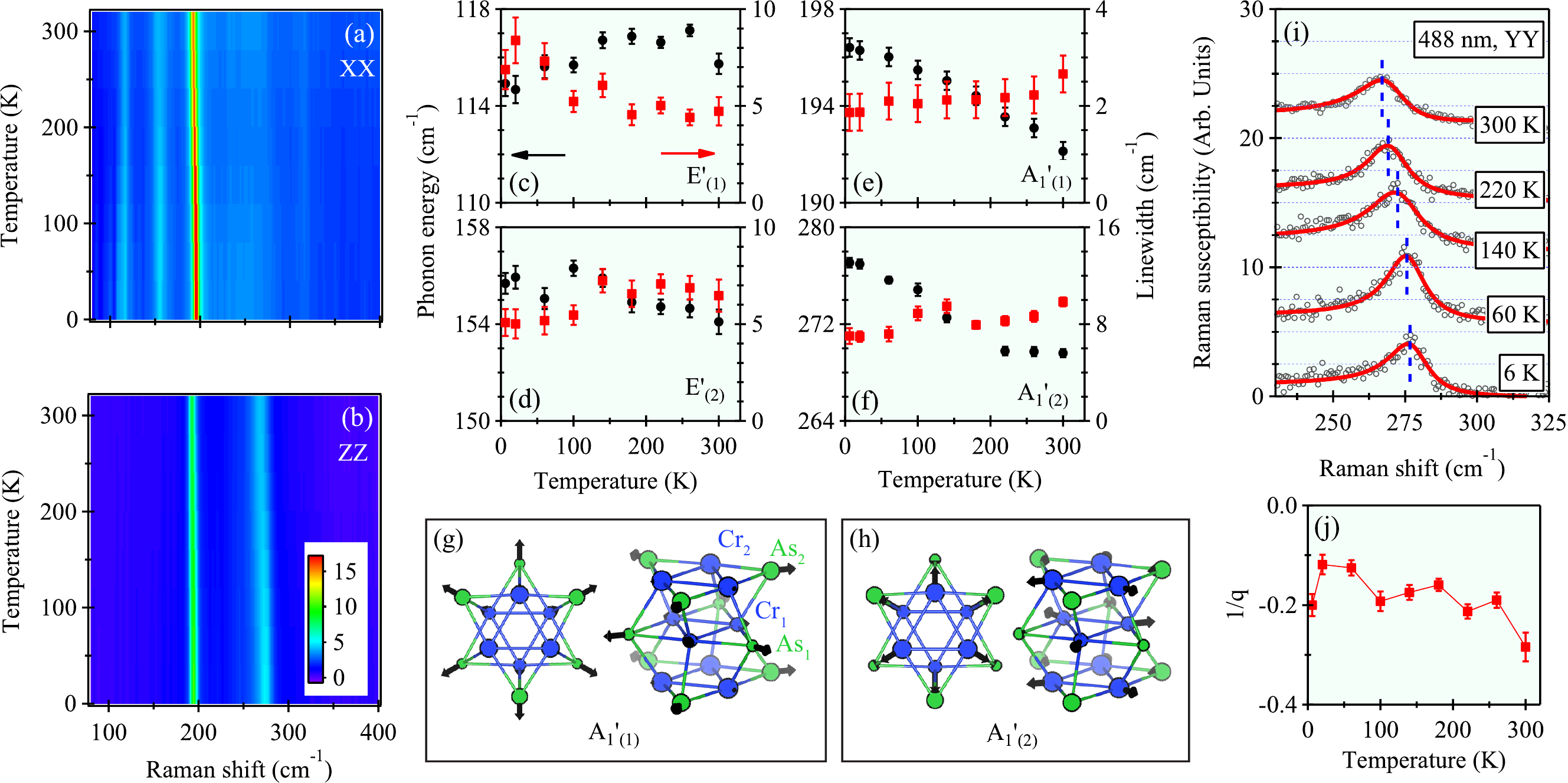}
\end{center}
 \caption{\label{Fig: 3} (Color online) (a) and (b) Intensity plot of the the temperature dependence of the Raman susceptibility under XX and ZZ polarizations, respectively. (c)-(f) Temperature evolution of the phonon energies and the line widths of the E$^\prime_{(1)}$, E$^\prime_{(2)}$, A$_1^\prime$$_{(1)}$ and A$_1^\prime$$_{(2)}$ peaks from the fitting of the data in (a) and (b) by eq.~\eqref{Fit}. (g) and (h) Atomic displacements in the CrAs tube of the  A$_1^\prime$$_{(1)}$ and A$_1^\prime$$_{(2)}$ modes, respectively, from top view and side view. (i) Zoom of the Raman response function and the Fano fitting curve of the A$_1^\prime$$_{(2)}$ phonon peak at representative temperatures. (j) Temperature dependence of the electron-phonon coupling strength 1/q by fitting data in (b) by eq.~\eqref{Fit}.}
\end{figure*}

The Raman response in the XX, ZZ and XZ polarization configurations recorded at room temperature under 488~nm laser excitation are plotted in Fig.~\ref{Fig: 2}. Similar results are obtained using the 514~nm laser excitation, as shown in the inset of Fig.~\ref{Fig: 2}. In the XX polarization we identify 5 Raman active modes at 115.7~cm$^{-1}$, 154.1~cm$^{-1}$, 192.1~cm$^{-1}$, 267.0~cm$^{-1}$ and 311.0~cm$^{-1}$. In order to separate the phonon modes with A$_1^\prime$ and E$^\prime$ symmetries, we also recorded a spectrum in the ZZ polarization. In the ZZ polarization, the 192.1~cm$^{-1}$ and the 267.0~cm$^{-1}$ peaks remain while the other three disappear. We thus conclude that the 192.1 cm$^{-1}$ (A$_1^\prime$$_{(1)}$) and the 267.0~cm$^{-1}$ (A$_1^\prime$$_{(2)}$) peaks have the A$_1^\prime$ symmetry whereas the 115.7~cm$^{-1}$ (E$^\prime_{(1)}$), 192.1~cm$^{-1} $(E$^\prime_{(2)}$) and  311.0~cm$^{-1} $(E$^\prime_{(3)}$) peaks are characterized by the E$^\prime$ symmetry. We also notice that the amplitude of A$_1^\prime$$_{(2)}$ increases in the ZZ polarization as compared to the XX polarization, in contrast to the amplitude of the A$_1^\prime$$_{(1)}$ mode which decreases in the XX polarized spectrum. This is due to the strong 1D character of \KCA, which leads to a strong anisotropy of the $a$ and $b$ terms in the Raman tensor. Despite their Raman activity under XZ polarization, no E$^{\prime\prime}$ symmetry mode is observed. 

Although the symmetries of the Raman modes observed can be identified experimentally with certainty, their assignments to precise vibrational modes is not always unique as some discrepancies are observed between the experimental phonon energies and the calculated values at the $\Gamma$ point, which are given in Table.~\ref{Table: Cal}. The calculated energies are also indicated by squares at the top of Fig. \ref{Fig: 1}. By taking into account the peak symmetries and their proximity in energy with the calculated values, we tentatively assign E$^\prime_{(1)}$, E$^\prime_{(2)}$ and E$^\prime_{(3)}$ to phonons No.~6, No. 11 and No. 20, respectively, while we assign A$_1^\prime$$_{(1)}$ and A$_1^\prime$$_{(2)}$ to phonons No. 14 and No. 18, respectively. As we will discuss below, one possible candidate to explain the discrepancies between experiments and calculations is the presence of a magneto-elastic coupling, which is neglected in the calculations.

In contrast to all the other peaks observed, the A$^\prime_1$$_{(2)}$ feature has a quite asymmetric lineshape. In Fig.~\ref{Fig: 2} we show that the Raman response in the XX and ZZ polarization configurations can be fitted by the equation:
\begin{align}
\label{Fit}
\chi^{\prime\prime}(\omega)
=\Sigma_i \,\, T_i\,L\,(\omega, \omega_i, \Gamma_i)+T_F\,F\,(\omega,\omega_F,\Gamma_F,q)
\end{align}
where $L\,(\omega, \omega_i, \Gamma_i)$ = $\frac{\Gamma_i}{(\omega-\omega_i)^2+\Gamma_i^2}$ is the Lorentz function characteristic of the normal phonon modes and $F\,(\omega,\omega_F,\Gamma_F,q)$=$\frac{1}{\Gamma_F q^2}\frac{[q+\alpha(\omega)]^2}{[1+\alpha(\omega)^2]}$ with $\alpha(\omega)=\frac{\omega-\omega_F}{\Gamma_F}$ is the Fano function ~\cite{Fano_PhysRev1961} well suited for the asymmetric peak at 267 cm$^{-1}$ (A$_1^\prime$$_{(2)}$). Here $T_i$($T_F$) is the amplitude of the $i^{\textrm{th}}$ phonon mode, while $\omega_i$($\omega_F$) and $\Gamma_i$($\Gamma_F$) are the bare phonon energy and line width, respectively. The $1/q$ value is commonly used in Raman spectroscopy as a measure of the electron-phonon coupling strength, such as in the cuprates \cite{Cooper_PRB1988,AMZhangCPB2013}, the Fe-based superconductors \cite{Chauviere_PRB2011,AMZhangCPB2013} and the Bi-based superconductors \cite{SFWu_PRB2014}.
For \KCA we get $1/q$ = - 0.29 $\pm$ 0.03 from the spectrum of ZZ polarization at 300~K. We estimate the electron-phonon coupling strength $\lambda=0.054$ from the Allen formula~\cite{Allen_PRB1972} using the A$_1^\prime$$_{(2)}$ phonon at 6~K, with a density-of-states at the Fermi level $N(0)=18$~eV$^{-1}$ per unit cell~\cite{XXWu_arxiV1503}. Although measurements below $T_c$ and a complete knowledge of all the phonon modes would be necessary to make a conclusive statement, the contribution of the A$_1^\prime$$_{(2)}$ mode to $\lambda=0.054$ is too small to afford for the $T_c$ of this material within the BCS picture.

We show in Figs.~\ref{Fig: 3}(a) and \ref{Fig: 3}(b) the temperature evolution from 6~K to 300~K of the intensity plot of the phonon spectra obtained with 488 nm laser excitation in the XX and ZZ polarization configurations. The spectra at all temperatures are qualitatively similar to the data recorded at 300 K. The phonon energies $\omega_i$ and line widths $\Gamma_i$ obtained by fitting the data to eq. \eqref{Fit} for E$^\prime_{(1)}$, E$^\prime_{(2)}$ and A$_1^\prime$$_{(1)}$ are plotted in Figs.~\ref{Fig: 3}(c)-\ref{Fig: 3}(f). The Raman response function of the unusual A$_1^\prime$$_{(2)}$ phonon under YY polarization and the fitting curve at representative temperatures are shown in Fig.~\ref{Fig: 3}(i). A distorted lineshape is observed at all temperatures. Indeed, Fig.~\ref{Fig: 3}(j) shows that the coupling constant 1/q extracted from the fits varies only a little with temperature. 

Although we cannot exclude that the Fano lineshape of the A$_1^\prime$$_{(2)}$ phonon is purely due to the electron-phonon coupling, many indications suggest that a coupling between the lattice and magnetic fluctuations is involved. As mentioned above, the experimental phonon energies are particularly difficult to reproduce from the calculations, possibly due to the influence of magnetic fluctuations. Additional information can be deduced from the spectral lineshape of the various Raman peaks. In Figs.~\ref{Fig: 3}(g) and \ref{Fig: 3}(h), we illustrate the vibration configuration of the sharp and symmetric A$_1^\prime$$_{(1)}$ phonon peak and that of the broader and asymmetric A$_1^\prime$$_{(2)}$ phonon mode, respectively. Unlike the A$_1^\prime$$_{(1)}$ mode that mainly involves in-phase co-planar vibrations of the As atoms, the A$_1^\prime$$_{(2)}$ mode involves mainly the in-phase co-planar vibrations of the Cr atoms. Consequently, the latter vibration mode has a much stronger impact on the intra-layer Cr-Cr bonding. Interestingly, a recent theoretical study indicates that the next-neighbor Cr-Cr exchange interaction is by far the strongest in this system~\cite{XXWu_CPL2015}, and the modulation of the Cr-Cr intra-layer distance should thus lead to the strongest magneto-elastic effect. The importance of the intra-layer Cr-Cr distance is also supported by the sensitivity of superconductivity to isovalent substitution of K with larger Cs~\cite{ZTTang_SCM2015} or Rb~\cite{ZTTang_PRB2015} atoms, and to applying hydrostatic or uniaxial pressure~\cite{ZWang_arXiv2015}. Even though our results do not allow us to comment directly on the unconventional nature of superconductivity in \KCA, they certainly suggest that the magnetic fluctuations reported in this system are coupled to the electronic structure \emph{via} lattice vibrations, which may be of crucial importance in elaborating models for superconductivity in this material.

We acknowledge P. Zhang, X.-X. Wu, S.-F. Wu and D. Chen for discussions. This work was supported by grants from MOST (2010CB923000, 2011CBA001000, 2011CBA00102, 2012CB821403 and 2013CB921703) and NSFC (11004232, 11034011/A0402, 11234014, 11274362 and 11474330) of China and by the Strategic Priority Research Program (B) of the Chinese Academy of Sciences, Grant No. XDB07020100.

\bibliography{KCA_bib}

\begin{thebibliography}{24}%
\makeatletter
\providecommand \@ifxundefined [1]{%
 \@ifx{#1\undefined}
}%
\providecommand \@ifnum [1]{%
 \ifnum #1\expandafter \@firstoftwo
 \else \expandafter \@secondoftwo
 \fi
}%
\providecommand \@ifx [1]{%
 \ifx #1\expandafter \@firstoftwo
 \else \expandafter \@secondoftwo
 \fi
}%
\providecommand \natexlab [1]{#1}%
\providecommand \enquote  [1]{``#1''}%
\providecommand \bibnamefont  [1]{#1}%
\providecommand \bibfnamefont [1]{#1}%
\providecommand \citenamefont [1]{#1}%
\providecommand \href@noop [0]{\@secondoftwo}%
\providecommand \href [0]{\begingroup \@sanitize@url \@href}%
\providecommand \@href[1]{\@@startlink{#1}\@@href}%
\providecommand \@@href[1]{\endgroup#1\@@endlink}%
\providecommand \@sanitize@url [0]{\catcode `\\12\catcode `\$12\catcode
  `\&12\catcode `\#12\catcode `\^12\catcode `\_12\catcode `\%12\relax}%
\providecommand \@@startlink[1]{}%
\providecommand \@@endlink[0]{}%
\providecommand \url  [0]{\begingroup\@sanitize@url \@url }%
\providecommand \@url [1]{\endgroup\@href {#1}{\urlprefix }}%
\providecommand \urlprefix  [0]{URL }%
\providecommand \Eprint [0]{\href }%
\providecommand \doibase [0]{http://dx.doi.org/}%
\providecommand \selectlanguage [0]{\@gobble}%
\providecommand \bibinfo  [0]{\@secondoftwo}%
\providecommand \bibfield  [0]{\@secondoftwo}%
\providecommand \translation [1]{[#1]}%
\providecommand \BibitemOpen [0]{}%
\providecommand \bibitemStop [0]{}%
\providecommand \bibitemNoStop [0]{.\EOS\space}%
\providecommand \EOS [0]{\spacefactor3000\relax}%
\providecommand \BibitemShut  [1]{\csname bibitem#1\endcsname}%
\let\auto@bib@innerbib\@empty
\bibitem [{\citenamefont {Fano}(1961)}]{Fano_PhysRev1961}%
  \BibitemOpen
  \bibfield  {author} {\bibinfo {author} {\bibfnamefont {U.}~\bibnamefont
  {Fano}},\ }\href {http://link.aps.org/doi/10.1103/PhysRev.124.1866}
  {\bibfield  {journal} {\bibinfo  {journal} {Phys. Rev.}\ }\textbf {\bibinfo
  {volume} {124}},\ \bibinfo {pages} {1866} (\bibinfo {year}
  {1961})}\BibitemShut {NoStop}%
\bibitem [{\citenamefont {Bao}\ \emph {et~al.}(2015)\citenamefont {Bao},
  \citenamefont {Liu}, \citenamefont {Ma}, \citenamefont {Meng}, \citenamefont
  {Tang}, \citenamefont {Sun}, \citenamefont {Zhai}, \citenamefont {Jiang},
  \citenamefont {Bai}, \citenamefont {Feng}, \citenamefont {Xu},\ and\
  \citenamefont {Cao}}]{JKBao_PRX2015}%
  \BibitemOpen
  \bibfield  {author} {\bibinfo {author} {\bibfnamefont {J.-K.}\ \bibnamefont
  {Bao}}, \bibinfo {author} {\bibfnamefont {J.-Y.}\ \bibnamefont {Liu}},
  \bibinfo {author} {\bibfnamefont {C.-W.}\ \bibnamefont {Ma}}, \bibinfo
  {author} {\bibfnamefont {Z.-H.}\ \bibnamefont {Meng}}, \bibinfo {author}
  {\bibfnamefont {Z.-T.}\ \bibnamefont {Tang}}, \bibinfo {author}
  {\bibfnamefont {Y.-L.}\ \bibnamefont {Sun}}, \bibinfo {author} {\bibfnamefont
  {H.-F.}\ \bibnamefont {Zhai}}, \bibinfo {author} {\bibfnamefont
  {H.}~\bibnamefont {Jiang}}, \bibinfo {author} {\bibfnamefont
  {H.}~\bibnamefont {Bai}}, \bibinfo {author} {\bibfnamefont {C.-M.}\
  \bibnamefont {Feng}}, \bibinfo {author} {\bibfnamefont {Z.-A.}\ \bibnamefont
  {Xu}}, \ and\ \bibinfo {author} {\bibfnamefont {G.-H.}\ \bibnamefont {Cao}},\
  }\href {http://link.aps.org/doi/10.1103/PhysRevX.5.011013} {\bibfield
  {journal} {\bibinfo  {journal} {Phys. Rev. X}\ }\textbf {\bibinfo {volume}
  {5}},\ \bibinfo {pages} {011013} (\bibinfo {year} {2015})}\BibitemShut
  {NoStop}%
\bibitem [{\citenamefont {Tang}\ \emph
  {et~al.}(2015{\natexlab{a}})\citenamefont {Tang}, \citenamefont {Bao},
  \citenamefont {Wang}, \citenamefont {Bai}, \citenamefont {Jiang},
  \citenamefont {Liu}, \citenamefont {Zhai}, \citenamefont {Feng},
  \citenamefont {Xu},\ and\ \citenamefont {Cao}}]{ZTTang_SCM2015}%
  \BibitemOpen
  \bibfield  {author} {\bibinfo {author} {\bibfnamefont {Z.-T.}\ \bibnamefont
  {Tang}}, \bibinfo {author} {\bibfnamefont {J.-K.}\ \bibnamefont {Bao}},
  \bibinfo {author} {\bibfnamefont {Z.}~\bibnamefont {Wang}}, \bibinfo {author}
  {\bibfnamefont {H.}~\bibnamefont {Bai}}, \bibinfo {author} {\bibfnamefont
  {H.}~\bibnamefont {Jiang}}, \bibinfo {author} {\bibfnamefont
  {Y.}~\bibnamefont {Liu}}, \bibinfo {author} {\bibfnamefont {H.-F.}\
  \bibnamefont {Zhai}}, \bibinfo {author} {\bibfnamefont {C.-M.}\ \bibnamefont
  {Feng}}, \bibinfo {author} {\bibfnamefont {Z.-A.}\ \bibnamefont {Xu}}, \ and\
  \bibinfo {author} {\bibfnamefont {G.-H.}\ \bibnamefont {Cao}},\ }\href
  {http://dx.doi.org/10.1007/s40843-015-0021-x} {\bibfield  {journal} {\bibinfo
   {journal} {Sci. China Mater.}\ }\textbf {\bibinfo {volume} {58}},\ \bibinfo
  {pages} {16} (\bibinfo {year} {2015}{\natexlab{a}})}\BibitemShut {NoStop}%
\bibitem [{\citenamefont {Tang}\ \emph
  {et~al.}(2015{\natexlab{b}})\citenamefont {Tang}, \citenamefont {Bao},
  \citenamefont {Liu}, \citenamefont {Sun}, \citenamefont {Ablimit},
  \citenamefont {Zhai}, \citenamefont {Jiang}, \citenamefont {Feng},
  \citenamefont {Xu},\ and\ \citenamefont {Cao}}]{ZTTang_PRB2015}%
  \BibitemOpen
  \bibfield  {author} {\bibinfo {author} {\bibfnamefont {Z.-T.}\ \bibnamefont
  {Tang}}, \bibinfo {author} {\bibfnamefont {J.-K.}\ \bibnamefont {Bao}},
  \bibinfo {author} {\bibfnamefont {Y.}~\bibnamefont {Liu}}, \bibinfo {author}
  {\bibfnamefont {Y.-L.}\ \bibnamefont {Sun}}, \bibinfo {author} {\bibfnamefont
  {A.}~\bibnamefont {Ablimit}}, \bibinfo {author} {\bibfnamefont {H.-F.}\
  \bibnamefont {Zhai}}, \bibinfo {author} {\bibfnamefont {H.}~\bibnamefont
  {Jiang}}, \bibinfo {author} {\bibfnamefont {C.-M.}\ \bibnamefont {Feng}},
  \bibinfo {author} {\bibfnamefont {Z.-A.}\ \bibnamefont {Xu}}, \ and\ \bibinfo
  {author} {\bibfnamefont {G.-H.}\ \bibnamefont {Cao}},\ }\href
  {http://link.aps.org/doi/10.1103/PhysRevB.91.020506} {\bibfield  {journal}
  {\bibinfo  {journal} {Phys. Rev. B}\ }\textbf {\bibinfo {volume} {91}},\
  \bibinfo {pages} {020506} (\bibinfo {year} {2015}{\natexlab{b}})}\BibitemShut
  {NoStop}%
\bibitem [{\citenamefont {Zhi}\ \emph {et~al.}(2015)\citenamefont {Zhi},
  \citenamefont {Imai}, \citenamefont {Ning}, \citenamefont {Bao},\ and\
  \citenamefont {Cao}}]{HZZhi_NMR2015}%
  \BibitemOpen
  \bibfield  {author} {\bibinfo {author} {\bibfnamefont {H.~Z.}\ \bibnamefont
  {Zhi}}, \bibinfo {author} {\bibfnamefont {T.}~\bibnamefont {Imai}}, \bibinfo
  {author} {\bibfnamefont {F.~L.}\ \bibnamefont {Ning}}, \bibinfo {author}
  {\bibfnamefont {J.-K.}\ \bibnamefont {Bao}}, \ and\ \bibinfo {author}
  {\bibfnamefont {G.-H.}\ \bibnamefont {Cao}},\ }\href
  {http://link.aps.org/doi/10.1103/PhysRevLett.114.147004} {\bibfield
  {journal} {\bibinfo  {journal} {Phys. Rev. Lett.}\ }\textbf {\bibinfo
  {volume} {114}},\ \bibinfo {pages} {147004} (\bibinfo {year}
  {2015})}\BibitemShut {NoStop}%
\bibitem [{\citenamefont {{Pang}}\ \emph {et~al.}()\citenamefont {{Pang}},
  \citenamefont {{Smidman}}, \citenamefont {{Jiang}}, \citenamefont {{Bao}},
  \citenamefont {{Weng}}, \citenamefont {{Wang}}, \citenamefont {{Jiao}},
  \citenamefont {{Zhang}}, \citenamefont {{Cao}},\ and\ \citenamefont
  {{Yuan}}}]{GMPang_arXiv2015}%
  \BibitemOpen
  \bibfield  {author} {\bibinfo {author} {\bibfnamefont {G.~M.}\ \bibnamefont
  {{Pang}}}, \bibinfo {author} {\bibfnamefont {M.}~\bibnamefont {{Smidman}}},
  \bibinfo {author} {\bibfnamefont {W.~B.}\ \bibnamefont {{Jiang}}}, \bibinfo
  {author} {\bibfnamefont {J.~K.}\ \bibnamefont {{Bao}}}, \bibinfo {author}
  {\bibfnamefont {Z.~F.}\ \bibnamefont {{Weng}}}, \bibinfo {author}
  {\bibfnamefont {Y.~F.}\ \bibnamefont {{Wang}}}, \bibinfo {author}
  {\bibfnamefont {L.}~\bibnamefont {{Jiao}}}, \bibinfo {author} {\bibfnamefont
  {J.~L.}\ \bibnamefont {{Zhang}}}, \bibinfo {author} {\bibfnamefont {G.~H.}\
  \bibnamefont {{Cao}}}, \ and\ \bibinfo {author} {\bibfnamefont {H.~Q.}\
  \bibnamefont {{Yuan}}},\ }\href
  {http://adsabs.harvard.edu/abs/2015arXiv150101880P} {\ }\Eprint
  {http://arxiv.org/abs/1501.01880} {arXiv:1501.01880} \BibitemShut {NoStop}%
\bibitem [{\citenamefont {{Wu}}\ \emph {et~al.}()\citenamefont {{Wu}},
  \citenamefont {{Yang}}, \citenamefont {{Le}}, \citenamefont {{Fan}},\ and\
  \citenamefont {{Hu}}}]{XXWu_arxiV1503}%
  \BibitemOpen
  \bibfield  {author} {\bibinfo {author} {\bibfnamefont {X.}~\bibnamefont
  {{Wu}}}, \bibinfo {author} {\bibfnamefont {F.}~\bibnamefont {{Yang}}},
  \bibinfo {author} {\bibfnamefont {C.}~\bibnamefont {{Le}}}, \bibinfo {author}
  {\bibfnamefont {H.}~\bibnamefont {{Fan}}}, \ and\ \bibinfo {author}
  {\bibfnamefont {J.}~\bibnamefont {{Hu}}},\ }\href
  {http://adsabs.harvard.edu/abs/2015arXiv150306707W} {\ }\Eprint
  {http://arxiv.org/abs/1503.06707} {arXiv:1503.06707} \BibitemShut {NoStop}%
\bibitem [{\citenamefont {Wu}\ \emph {et~al.}(2014{\natexlab{a}})\citenamefont
  {Wu}, \citenamefont {Cheng}, \citenamefont {Matsubayashi}, \citenamefont
  {Kong}, \citenamefont {Lin}, \citenamefont {Jin}, \citenamefont {Wang},
  \citenamefont {Uwatoko},\ and\ \citenamefont {Luo}}]{W_Wu5ncomm}%
  \BibitemOpen
  \bibfield  {author} {\bibinfo {author} {\bibfnamefont {W.}~\bibnamefont
  {Wu}}, \bibinfo {author} {\bibfnamefont {J.}~\bibnamefont {Cheng}}, \bibinfo
  {author} {\bibfnamefont {K.}~\bibnamefont {Matsubayashi}}, \bibinfo {author}
  {\bibfnamefont {P.}~\bibnamefont {Kong}}, \bibinfo {author} {\bibfnamefont
  {F.}~\bibnamefont {Lin}}, \bibinfo {author} {\bibfnamefont {C.}~\bibnamefont
  {Jin}}, \bibinfo {author} {\bibfnamefont {N.}~\bibnamefont {Wang}}, \bibinfo
  {author} {\bibfnamefont {Y.}~\bibnamefont {Uwatoko}}, \ and\ \bibinfo
  {author} {\bibfnamefont {J.}~\bibnamefont {Luo}},\ }\href
  {http://dx.doi.org/10.1038/ncomms6508} {\bibfield  {journal} {\bibinfo
  {journal} {Nat. Commun.}\ }\textbf {\bibinfo {volume} {5}},\ \bibinfo {pages}
  {5508} (\bibinfo {year} {2014}{\natexlab{a}})}\BibitemShut {NoStop}%
\bibitem [{\citenamefont {{Jiang}}\ \emph {et~al.}()\citenamefont {{Jiang}},
  \citenamefont {{Cao}},\ and\ \citenamefont {{Cao}}}]{HaoJiang_arXiv2014}%
  \BibitemOpen
  \bibfield  {author} {\bibinfo {author} {\bibfnamefont {H.}~\bibnamefont
  {{Jiang}}}, \bibinfo {author} {\bibfnamefont {G.}~\bibnamefont {{Cao}}}, \
  and\ \bibinfo {author} {\bibfnamefont {C.}~\bibnamefont {{Cao}}},\ }\href
  {http://adsabs.harvard.edu/abs/2014arXiv1412.1309J} {\ }\Eprint
  {http://arxiv.org/abs/1412.1309} {arXiv:1412.1309} \BibitemShut {NoStop}%
\bibitem [{\citenamefont {Wu}\ \emph {et~al.}(2015)\citenamefont {Wu},
  \citenamefont {Le}, \citenamefont {Yuan}, \citenamefont {Fan},\ and\
  \citenamefont {Hu}}]{XXWu_CPL2015}%
  \BibitemOpen
  \bibfield  {author} {\bibinfo {author} {\bibfnamefont {X.-X.}\ \bibnamefont
  {Wu}}, \bibinfo {author} {\bibfnamefont {C.-C.}\ \bibnamefont {Le}}, \bibinfo
  {author} {\bibfnamefont {J.}~\bibnamefont {Yuan}}, \bibinfo {author}
  {\bibfnamefont {H.}~\bibnamefont {Fan}}, \ and\ \bibinfo {author}
  {\bibfnamefont {J.-P.}\ \bibnamefont {Hu}},\ }\href
  {http://cpl.iphy.ac.cn/EN/abstract/article_64390.shtml} {\bibfield  {journal}
  {\bibinfo  {journal} {Chi. Phys. Lett.}\ }\textbf {\bibinfo {volume} {32}},\
  \bibinfo {eid} {57401} (\bibinfo {year} {2015})}\BibitemShut {NoStop}%
\bibitem [{\citenamefont {Kong}\ \emph {et~al.}(2015)\citenamefont {Kong},
  \citenamefont {Bud'ko},\ and\ \citenamefont {Canfield}}]{KTai_PRB2015}%
  \BibitemOpen
  \bibfield  {author} {\bibinfo {author} {\bibfnamefont {T.}~\bibnamefont
  {Kong}}, \bibinfo {author} {\bibfnamefont {S.~L.}\ \bibnamefont {Bud'ko}}, \
  and\ \bibinfo {author} {\bibfnamefont {P.~C.}\ \bibnamefont {Canfield}},\
  }\href {http://link.aps.org/doi/10.1103/PhysRevB.91.020507} {\bibfield
  {journal} {\bibinfo  {journal} {Phys. Rev. B}\ }\textbf {\bibinfo {volume}
  {91}},\ \bibinfo {pages} {020507} (\bibinfo {year} {2015})}\BibitemShut
  {NoStop}%
\bibitem [{\citenamefont {Kresse}\ and\ \citenamefont
  {Hafner}(1993)}]{Kresse1993}%
  \BibitemOpen
  \bibfield  {author} {\bibinfo {author} {\bibfnamefont {G.}~\bibnamefont
  {Kresse}}\ and\ \bibinfo {author} {\bibfnamefont {J.}~\bibnamefont
  {Hafner}},\ }\href {http://link.aps.org/doi/10.1103/PhysRevB.47.558}
  {\bibfield  {journal} {\bibinfo  {journal} {Phys. Rev. B}\ }\textbf {\bibinfo
  {volume} {47}},\ \bibinfo {pages} {558} (\bibinfo {year} {1993})}\BibitemShut
  {NoStop}%
\bibitem [{\citenamefont {Kresse}\ and\ \citenamefont
  {Furthm\"uller}(1996{\natexlab{a}})}]{Kresse1996}%
  \BibitemOpen
  \bibfield  {author} {\bibinfo {author} {\bibfnamefont {G.}~\bibnamefont
  {Kresse}}\ and\ \bibinfo {author} {\bibfnamefont {J.}~\bibnamefont
  {Furthm\"uller}},\ }\href
  {http://www.sciencedirect.com/science/article/pii/0927025696000080}
  {\bibfield  {journal} {\bibinfo  {journal} {Comput. Mater. Sci.}\ }\textbf
  {\bibinfo {volume} {6}},\ \bibinfo {pages} {15 } (\bibinfo {year}
  {1996}{\natexlab{a}})}\BibitemShut {NoStop}%
\bibitem [{\citenamefont {Kresse}\ and\ \citenamefont
  {Furthm\"uller}(1996{\natexlab{b}})}]{Kresse1996B}%
  \BibitemOpen
  \bibfield  {author} {\bibinfo {author} {\bibfnamefont {G.}~\bibnamefont
  {Kresse}}\ and\ \bibinfo {author} {\bibfnamefont {J.}~\bibnamefont
  {Furthm\"uller}},\ }\href {http://link.aps.org/doi/10.1103/PhysRevB.54.11169}
  {\bibfield  {journal} {\bibinfo  {journal} {Phys. Rev. B}\ }\textbf {\bibinfo
  {volume} {54}},\ \bibinfo {pages} {11169} (\bibinfo {year}
  {1996}{\natexlab{b}})}\BibitemShut {NoStop}%
\bibitem [{\citenamefont {Perdew}\ \emph {et~al.}(1996)\citenamefont {Perdew},
  \citenamefont {Burke},\ and\ \citenamefont {Ernzerhof}}]{Perdew1996}%
  \BibitemOpen
  \bibfield  {author} {\bibinfo {author} {\bibfnamefont {J.~P.}\ \bibnamefont
  {Perdew}}, \bibinfo {author} {\bibfnamefont {K.}~\bibnamefont {Burke}}, \
  and\ \bibinfo {author} {\bibfnamefont {M.}~\bibnamefont {Ernzerhof}},\ }\href
  {http://link.aps.org/doi/10.1103/PhysRevLett.77.3865} {\bibfield  {journal}
  {\bibinfo  {journal} {Phys. Rev. Lett.}\ }\textbf {\bibinfo {volume} {77}},\
  \bibinfo {pages} {3865} (\bibinfo {year} {1996})}\BibitemShut {NoStop}%
\bibitem [{\citenamefont {Baroni}\ \emph {et~al.}(1987)\citenamefont {Baroni},
  \citenamefont {Giannozzi},\ and\ \citenamefont {Testa}}]{Baroni}%
  \BibitemOpen
  \bibfield  {author} {\bibinfo {author} {\bibfnamefont {S.}~\bibnamefont
  {Baroni}}, \bibinfo {author} {\bibfnamefont {P.}~\bibnamefont {Giannozzi}}, \
  and\ \bibinfo {author} {\bibfnamefont {A.}~\bibnamefont {Testa}},\ }\href
  {http://link.aps.org/doi/10.1103/PhysRevLett.58.1861} {\bibfield  {journal}
  {\bibinfo  {journal} {Phys. Rev. Lett.}\ }\textbf {\bibinfo {volume} {58}},\
  \bibinfo {pages} {1861} (\bibinfo {year} {1987})}\BibitemShut {NoStop}%
\bibitem [{\citenamefont {Togo}\ \emph {et~al.}(2008)\citenamefont {Togo},
  \citenamefont {Oba},\ and\ \citenamefont {Tanaka}}]{phonopy}%
  \BibitemOpen
  \bibfield  {author} {\bibinfo {author} {\bibfnamefont {A.}~\bibnamefont
  {Togo}}, \bibinfo {author} {\bibfnamefont {F.}~\bibnamefont {Oba}}, \ and\
  \bibinfo {author} {\bibfnamefont {I.}~\bibnamefont {Tanaka}},\ }\href
  {http://link.aps.org/doi/10.1103/PhysRevB.78.134106} {\bibfield  {journal}
  {\bibinfo  {journal} {Phys. Rev. B}\ }\textbf {\bibinfo {volume} {78}},\
  \bibinfo {pages} {134106} (\bibinfo {year} {2008})}\BibitemShut {NoStop}%
\bibitem [{\citenamefont {Kroumova}\ \emph {et~al.}(2003)\citenamefont
  {Kroumova}, \citenamefont {Aroyo}, \citenamefont {Perez-Mato}, \citenamefont
  {Kirov}, \citenamefont {Capillas}, \citenamefont {Ivantchev},\ and\
  \citenamefont {Wondratschek}}]{Bilbao_1}%
  \BibitemOpen
  \bibfield  {author} {\bibinfo {author} {\bibfnamefont {E.}~\bibnamefont
  {Kroumova}}, \bibinfo {author} {\bibfnamefont {M.}~\bibnamefont {Aroyo}},
  \bibinfo {author} {\bibfnamefont {J.}~\bibnamefont {Perez-Mato}}, \bibinfo
  {author} {\bibfnamefont {A.}~\bibnamefont {Kirov}}, \bibinfo {author}
  {\bibfnamefont {C.}~\bibnamefont {Capillas}}, \bibinfo {author}
  {\bibfnamefont {S.}~\bibnamefont {Ivantchev}}, \ and\ \bibinfo {author}
  {\bibfnamefont {H.}~\bibnamefont {Wondratschek}},\ }\href
  {http://dx.doi.org/10.1080/0141159031000076110} {\bibfield  {journal}
  {\bibinfo  {journal} {Phase Transit.}\ }\textbf {\bibinfo {volume} {76}},\
  \bibinfo {pages} {155} (\bibinfo {year} {2003})}\BibitemShut {NoStop}%
\bibitem [{\citenamefont {Cooper}\ \emph {et~al.}(1988)\citenamefont {Cooper},
  \citenamefont {Klein}, \citenamefont {Pazol}, \citenamefont {Rice},\ and\
  \citenamefont {Ginsberg}}]{Cooper_PRB1988}%
  \BibitemOpen
  \bibfield  {author} {\bibinfo {author} {\bibfnamefont {S.~L.}\ \bibnamefont
  {Cooper}}, \bibinfo {author} {\bibfnamefont {M.~V.}\ \bibnamefont {Klein}},
  \bibinfo {author} {\bibfnamefont {B.~G.}\ \bibnamefont {Pazol}}, \bibinfo
  {author} {\bibfnamefont {J.~P.}\ \bibnamefont {Rice}}, \ and\ \bibinfo
  {author} {\bibfnamefont {D.~M.}\ \bibnamefont {Ginsberg}},\ }\href
  {http://link.aps.org/doi/10.1103/PhysRevB.37.5920} {\bibfield  {journal}
  {\bibinfo  {journal} {Phys. Rev. B}\ }\textbf {\bibinfo {volume} {37}},\
  \bibinfo {pages} {5920} (\bibinfo {year} {1988})}\BibitemShut {NoStop}%
\bibitem [{\citenamefont {Zhang}\ and\ \citenamefont
  {Zhang}(2013)}]{AMZhangCPB2013}%
  \BibitemOpen
  \bibfield  {author} {\bibinfo {author} {\bibfnamefont {A.-M.}\ \bibnamefont
  {Zhang}}\ and\ \bibinfo {author} {\bibfnamefont {Q.-M.}\ \bibnamefont
  {Zhang}},\ }\href {http://cpb.iphy.ac.cn/EN/abstract/article_54994.shtml}
  {\bibfield  {journal} {\bibinfo  {journal} {Chin. Phys. B}\ }\textbf
  {\bibinfo {volume} {22}},\ \bibinfo {eid} {87103} (\bibinfo {year}
  {2013})}\BibitemShut {NoStop}%
\bibitem [{\citenamefont {Chauvi\`ere}\ \emph {et~al.}(2011)\citenamefont
  {Chauvi\`ere}, \citenamefont {Gallais}, \citenamefont {Cazayous},
  \citenamefont {M\'easson}, \citenamefont {Sacuto}, \citenamefont {Colson},\
  and\ \citenamefont {Forget}}]{Chauviere_PRB2011}%
  \BibitemOpen
  \bibfield  {author} {\bibinfo {author} {\bibfnamefont {L.}~\bibnamefont
  {Chauvi\`ere}}, \bibinfo {author} {\bibfnamefont {Y.}~\bibnamefont
  {Gallais}}, \bibinfo {author} {\bibfnamefont {M.}~\bibnamefont {Cazayous}},
  \bibinfo {author} {\bibfnamefont {M.~A.}\ \bibnamefont {M\'easson}}, \bibinfo
  {author} {\bibfnamefont {A.}~\bibnamefont {Sacuto}}, \bibinfo {author}
  {\bibfnamefont {D.}~\bibnamefont {Colson}}, \ and\ \bibinfo {author}
  {\bibfnamefont {A.}~\bibnamefont {Forget}},\ }\href
  {http://link.aps.org/doi/10.1103/PhysRevB.84.104508} {\bibfield  {journal}
  {\bibinfo  {journal} {Phys. Rev. B}\ }\textbf {\bibinfo {volume} {84}},\
  \bibinfo {pages} {104508} (\bibinfo {year} {2011})}\BibitemShut {NoStop}%
\bibitem [{\citenamefont {Wu}\ \emph {et~al.}(2014{\natexlab{b}})\citenamefont
  {Wu}, \citenamefont {Richard}, \citenamefont {Wang}, \citenamefont {Lian},
  \citenamefont {Nie}, \citenamefont {Wang}, \citenamefont {Wang},\ and\
  \citenamefont {Ding}}]{SFWu_PRB2014}%
  \BibitemOpen
  \bibfield  {author} {\bibinfo {author} {\bibfnamefont {S.~F.}\ \bibnamefont
  {Wu}}, \bibinfo {author} {\bibfnamefont {P.}~\bibnamefont {Richard}},
  \bibinfo {author} {\bibfnamefont {X.~B.}\ \bibnamefont {Wang}}, \bibinfo
  {author} {\bibfnamefont {C.~S.}\ \bibnamefont {Lian}}, \bibinfo {author}
  {\bibfnamefont {S.~M.}\ \bibnamefont {Nie}}, \bibinfo {author} {\bibfnamefont
  {J.~T.}\ \bibnamefont {Wang}}, \bibinfo {author} {\bibfnamefont {N.~L.}\
  \bibnamefont {Wang}}, \ and\ \bibinfo {author} {\bibfnamefont
  {H.}~\bibnamefont {Ding}},\ }\href
  {http://link.aps.org/doi/10.1103/PhysRevB.90.054519} {\bibfield  {journal}
  {\bibinfo  {journal} {Phys. Rev. B}\ }\textbf {\bibinfo {volume} {90}},\
  \bibinfo {pages} {054519} (\bibinfo {year} {2014}{\natexlab{b}})}\BibitemShut
  {NoStop}%
\bibitem [{\citenamefont {Allen}(1972)}]{Allen_PRB1972}%
  \BibitemOpen
  \bibfield  {author} {\bibinfo {author} {\bibfnamefont {P.~B.}\ \bibnamefont
  {Allen}},\ }\href {http://link.aps.org/doi/10.1103/PhysRevB.6.2577}
  {\bibfield  {journal} {\bibinfo  {journal} {Phys. Rev. B}\ }\textbf {\bibinfo
  {volume} {6}},\ \bibinfo {pages} {2577} (\bibinfo {year} {1972})}\BibitemShut
  {NoStop}%
\bibitem [{\citenamefont {{Wang}}\ \emph {et~al.}()\citenamefont {{Wang}},
  \citenamefont {{Sidorov}}, \citenamefont {{Wu}}, \citenamefont {{Bao}},
  \citenamefont {{Tang}}, \citenamefont {{Yi}}, \citenamefont {{Guo}},
  \citenamefont {{Zhou}}, \citenamefont {{Zhang}}, \citenamefont {{Zhang}},
  \citenamefont {{Cao}}, \citenamefont {{Sun}},\ and\ \citenamefont
  {{Zhao}}}]{ZWang_arXiv2015}%
  \BibitemOpen
  \bibfield  {author} {\bibinfo {author} {\bibfnamefont {Z.}~\bibnamefont
  {{Wang}}}, \bibinfo {author} {\bibfnamefont {V.}~\bibnamefont {{Sidorov}}},
  \bibinfo {author} {\bibfnamefont {Q.}~\bibnamefont {{Wu}}}, \bibinfo {author}
  {\bibfnamefont {J.}~\bibnamefont {{Bao}}}, \bibinfo {author} {\bibfnamefont
  {Z.}~\bibnamefont {{Tang}}}, \bibinfo {author} {\bibfnamefont
  {W.}~\bibnamefont {{Yi}}}, \bibinfo {author} {\bibfnamefont {J.}~\bibnamefont
  {{Guo}}}, \bibinfo {author} {\bibfnamefont {Y.}~\bibnamefont {{Zhou}}},
  \bibinfo {author} {\bibfnamefont {S.}~\bibnamefont {{Zhang}}}, \bibinfo
  {author} {\bibfnamefont {C.}~\bibnamefont {{Zhang}}}, \bibinfo {author}
  {\bibfnamefont {G.}~\bibnamefont {{Cao}}}, \bibinfo {author} {\bibfnamefont
  {L.}~\bibnamefont {{Sun}}}, \ and\ \bibinfo {author} {\bibfnamefont
  {Z.}~\bibnamefont {{Zhao}}},\ }\href
  {http://adsabs.harvard.edu/abs/2015arXiv150204304W} {\ }\Eprint
  {http://arxiv.org/abs/1502.04304} {arXiv:1502.04304} \BibitemShut {NoStop}%
\end{thebibliography}%

\end{document}